# Entanglement-induced Kondo screening in atomic spin chains

Deung-Jang Choi[1,2,§], Roberto Robles[3], Shichao Yan[1,2], Jacob A. J. Burgess[1,2], Steffen Rolf-Pissarczyk[1,2], Jean-Pierre Gauyacq[4], Nicolás Lorente[5,6], Markus Ternes[2], Sebastian Loth[1,2,*]

[1]   *Max Planck Institute for the Structure and Dynamics of Matter, Luruper Chaussee 149, 22761 Hamburg, Germany*
[2]   *Max Planck Institute for Solid State Research, Heisenbergstr. 1, 70569 Stuttgart, Germany*
[3]   *ICN2 - Institut Catala de Nanociencia i Nanotecnologia, Campus UAB, 08193 Bellaterra (Barcelona), Spain*
[4]   *Institut des Sciences Moléculaires d'Orsay, ISMO, Unité Mixte de Recherches CNRS-Université Paris-Sud, UMR 8214,Bâtiment 351, Université Paris-Sud, 91405 Orsay Cedex, France*
[5]   *Centro de Física de Materiales, CFM/MPC (CSIC-UPV/EHU), Paseo Manuel de Lardizabal 5, 20018 Donostia-San Sebastián, Spain*
[6]   *Donostia International Physics Center (DIPC), Paseo Manuel de Lardizabal 4, 20018 Donostia-San Sebastián, Spain*
[§]   *Present address: CIC nanoGUNE, Tolosa Hiribidea 78, Donostia-San Sebastián 20018, Spain*
[*]   *corresponding author: sebastian.loth@mpsd.mpg.de*

**Quantum entanglement permeates the complex ground states of correlated electron materials defying single-particle descriptions. Coupled magnetic atoms have potential as model systems for entanglement in condensed matter[1] giving the opportunity to create artificial many-body states which can be controlled by tuning the underlying interactions [2]. They provide an avenue to unravel the complexities of correlated-electron materials[3]. Here we use low temperature scanning tunnelling microscopy (STM) and atomic manipulation to tune entanglement in chains of magnetic atoms. We find that a Kondo singlet state[4] can emerge from this entanglement. The many-electron Kondo state is based on the screening of the entangled spin ground state of the chain by substrate electrons and can be engineered to envelop at least ten magnetic atoms. The concomitant Kondo resonance measured in the differential conductance enables the electric read-out of entanglement. By tuning composition and coupling strength within atomic chains it is possible to create model spin chains with defined entanglement. This lays the foundation for a new class of experiments to construct exotic correlated-electron materials atom by atom.**

Coupled spins constitute the paragon of many-body entanglement studies[5]. An extensive body of theoretical work explores the rich range of new states of matter that can be created from the mutual interaction of spins in chains and lattices. Solutions range from classical-like states of the Ising model to the fully entangled states of the Heisenberg model[6] and encompass exotic effects such as fractionalization in Haldane chains[7,8], Luttinger liquid behaviour[9,10] or edge states in topological superconductors[11,12]. Long-range-entangled spin chains have even been proposed as means for quantum information transport[13].

Experimental realizations of entangled spin chains include anisotropic bulk materials[14], cold atom lattices[15] and artificial nanostructures on surfaces[12,16-18] which show that entanglement can persist in the presence of decoherence caused by the environment.  The possibility to intentionally create new correlated-electron states from entanglement in solid state systems has remained elusive thus far.

Here we engineer Kondo singlet states from quantum entanglement in 1D chains of magnetic atoms on a surface. The degree of entanglement and consequently the properties of the Kondo singlet state can be controlled by composition, coupling strength and length of the chains. We find that the coherent spin-flip scattering of substrate electrons required for the formation of the Kondo state emerges from entanglement of the spins within the chain allowing the creation of a new Kondo phase that is not based on the screening of individual atoms.

Experimentally, the Kondo singlet state itself manifests in spectra of the differential conductance, $dI/dV(V)$, as a prominent resonance located at zero bias[19]. Kondo resonances of individual spins have been observed in a variety of atoms and molecules on surfaces[20-23]. In this work we use a scanning tunnelling microscope to assemble disparate magnetic atoms into composite chains of up to 10 atoms with precisely defined geometry and elemental composition (Fig. 1a). Specifically, we use Fe atoms (with spin magnitude $S_{Fe} = 2$) and Mn atoms ($S_{Mn} = 5/2$)[24], and place them on the Cu binding sites of a $Cu_2N/Cu(100)$ surface (See Methods for details). When Fe and Mn are assembled into dimers with 0.36 nm or less separation, a sharp Kondo resonance appears in $dI/dV(V)$ at zero bias (Fig. 2b). Measurements of the temperature-dependent broadening of the conductance peak confirm its correlated nature (Supplementary Figure S1).

The amplitude of the Kondo resonance in the MnFe dimers is comparable to resonances observed for individual magnetic impurities on the $Cu_2N$ surface [21]. Unlike in the single-impurity case, the Kondo screening of the MnFe dimers cannot be caused by screening of the individual atoms. Due to their easy axis anisotropy, neither Fe nor Mn form a Kondo state at the temperature of our experiment, $T = 0.5$ K [25]. In addition, MnFe dimers with large separation and consequently weak spin-spin coupling show no Kondo resonance (Fig. 2b). Hence, the emergence of a Kondo singlet state in MnFe dimers must be related to collective states of the spin-coupled atoms.

To gain insight into this process we consider spin-coupling between Fe and Mn by Heisenberg exchange interaction using a simplified spin Hamiltonian model,

$$\hat{H}_1 = J\vec{S}_{Fe}\vec{S}_{Mn} - D\hat{S}_{Fe,z}^2,$$

and treat Kondo scattering up to 3rd order in the impurity spin (see Methods and Supplementary Figure S2 for details). This method has been demonstrated to capture the essential properties of Kondo resonances both in atomic and molecular systems [22]. The experimental $dI/dV(V)$ spectra which reflect the spin excitations and Kondo resonance of the dimers are reproduced well by the calculations (Fig. 2c) when accounting for an antiferromagnetic isotropic Heisenberg coupling, $J$, between Fe and Mn and easy-axis magnetic anisotropy, $D$, of the Fe atom.

The MnFe dimers feature a doubly degenerate ground state with $m_T = \pm 1/2$ ($m_T$ is component of the total spin, $\vec{S}_T = \vec{S}_{Fe} + \vec{S}_{Mn}$, projected onto the anisotropy axis of the Fe atom) and two low-energy excitations to doublets with $m_T = \{\pm 1/2, \pm 3/2\}$ (Fig. 1b). The spin excitations appear as steps in the $dI/dV(V)$ spectra at voltages $\pm E_1/e$ and $\pm E_2/e$ and permit extracting the Heisenberg coupling, $J$, and magnetic anisotropy constant, $D$ (Fig. 2ab). By plotting the ratio of the measured spin excitation energies, $(E_2+E_1)/(E_2-E_1)$, we find that the spin states of the MnFe dimer are completely characterized by the ratio $J/|D|$ (Fig. 2e). We can therefore use our model Hamiltonian to inspect the spin ground states for dimers with different coupling strengths.

The relation between entanglement and Kondo screening becomes apparent by comparing dimers with varying coupling strength, i.e., varying $J/|D|$ (Fig. 1c). For antiferromagnetically coupled dimers with $J/|D| \ll 1$, spin-spin coupling is small compared to the magnetic anisotropy (Fig. 2d dimer A) and the atomic spins align with the anisotropy axis. The two degenerate spin ground states become product states of the individual spins with $m_{Fe} = -2$, $m_{Mn} = +5/2$, and $m_{Fe} = +2$, $m_{Mn} = -5/2$ with vanishing entanglement ($m_{Fe}$ and $m_{Mn}$ are the magnetic quantum numbers of the individual atom spins $\vec{S}_{Fe}$ and $\vec{S}_{Mn}$). The absence of a Kondo resonance in the measured and calculated d$I$/d$V$($V$) spectra shows that these Néel-like states are not amenable to Kondo screening.

By comparison, strongly coupled antiferromagnetic dimers with $J/|D| \gg 1$ exhibit prominent Kondo resonances. Their ground states are superpositions of all spin states with $m_{Fe} + m_{Mn} = +^1/_2$ or $-^1/_2$ (Fig. 2d dimers B, C). The dimer behaves as a single global angular momentum (macrospin) of magnitude ½ (Fig. 2f,g). Consequently, a single-electron interaction at any atom of the chain can flip between the two degenerate ground states of the dimer and permit a Kondo singlet as the new ground state. By evaluating the entanglement entropy for these ground states (see Methods for details) we find that the complex spin superpositions are caused by entanglement of the Fe and Mn spins (Fig. 2f), linking the emergence of the Kondo resonance in the MnFe dimers to their degree of entanglement.

All other configurations with intermediate $J/|D|$ have partial entanglement of the spins and the observed strength of the Kondo resonance is reduced (Fig. 2g). This behaviour is consistent with a reduction in Kondo screening efficiency caused by an increasing weight of spin components with total spin in excess of ½ in the ground states as can be validated by the expectation value of the total spin, $\langle S_T^2 \rangle$. For the maximally entangled ground state $\langle S_T^2 \rangle$ is $(S_{Mn}-S_{Fe})(S_{Mn}-S_{Fe}+1) = 0.75$ and grows gradually to $S_{Fe}(S_{Fe}+1)+(S_{Mn})(S_{Mn}+1) = 14.75$ when entanglement is absent (Fig. 2g). Because the high-spin components of the superposition cannot be screened by single-electron interactions, a weaker entanglement drives the MnFe system into an underscreened Kondo regime[26]. As screening decreases the Kondo resonance decreases in strength as well [27] (See Supplementary Section S1 for details). Hence, the presence of a Kondo resonance in the MnFe dimers directly indicates strong quantum entanglement.

To test whether the Kondo singlet state can be extended to a larger number of atoms we assemble successively longer spin chains by adding Mn atoms to the MnFe dimer one by one (Fig. 3a,b). Surprisingly, all chains with odd number of Mn atoms, Mn$_x$Fe (x = 3, 5, 7, 9) feature a prominent Kondo resonance (Fig. 3d).

It is important to draw a distinction between the effect of quantum entanglement of the spins and orbital hybridization. Density functional theory (DFT) calculations find that all of the magnetic moment of the chain is localized in the d-orbitals of the Fe and Mn atoms (Fig. 3c). No spin-polarized molecular states were found that could participate in the Kondo screening (Fig. 3c and Supplementary Figure S3) indicating that – even up to the longest chains – the Kondo resonance is caused by quantum entanglement of all atoms in the chain. This conclusion is corroborated by the observed disappearance of the Kondo resonance upon adding one Mn atom to each chain (see Supplementary Figure S4 for spectra of Mn$_x$Fe (x = 2, 4, 6, 8, 10) ).

We identify the spin ground state responsible for Kondo screening in longer chains Mn$_x$Fe by extending our model Hamiltonian with nearest-neighbor exchange interaction for the additional Mn atoms[18],

$$\hat{H}_x = J\vec{S}_{Fe}\vec{S}_{Mn} - D\hat{S}_{Fe,z}^2 + \sum_{i=2..x} J'\vec{S}_{Mn,i-1}\vec{S}_{Mn,i} \ .$$

We find that the whole series from MnFe to Mn$_9$Fe can be fit with one set of parameters, $D = 4.0$ meV $\pm 0.4$ meV, $J = 13.2$ meV $\pm 1.0$ meV, $J' = 4.2$ meV $\pm 0.8$ meV ($J'$ is the coupling strength between Mn atoms), permitting quantitative comparisons among the chains (Supplementary Figure S5). All chains feature doubly degenerate and strongly entangled ground states with $m_T = \pm^1/_2$. The expectation value of the total spin, $\langle S_T^2 \rangle$, ranges between 0.9 for Mn$_1$Fe and 1.7 for Mn$_9$Fe indicating that the ground states of all chains are close to maximally entangled. This makes it possible for a single-electron interaction to screen the chain's magnetic moment and to form a spatially-extended Kondo singlet state.

Conductance spectra acquired along the $Mn_xFe$ chains, $dI/dV(V,d)$, show that the Kondo resonance is delocalized within the chain, as may be expected from a macrospin ground state. Remarkably, it is not distributed uniformly. The strongest resonance is at the Mn atom furthest from the Fe atom (Fig. 3d). The spatial profile of the Kondo resonance (Fig. 4a) shows that the resonance is present over all atoms in short chains $Mn_xFe$ (x = 1, 3) whereas it drops off at approximately 1.5 nm distance from the edge Mn atom in the longer chains (x = 5, 7, 9). This emphasizes that the edge Mn atom, even for the longest chain ($Mn_9Fe$), is entangled with the Fe atom on the other end.

The spatial profiles of the Kondo resonance show that the maximum amplitude increases approximately linearly with chain length and for $Mn_9Fe$ reaches up to 80% of the amplitude of the spin-excitation-induced conductance steps (Fig. 4a, inset). This behaviour is consistent with an increased efficiency of electron-spin scattering between the ground state doublet by substrate electrons due to the increased number of scattering sites in longer chains. In addition, the Fe atom's magnetic anisotropy, $D$, becomes effectively diluted with increasing number of Mn atoms allowing for a more efficient Kondo screening.

The presence of magnetic anisotropy becomes evident with application of a magnetic field along the high symmetry directions of the $Mn_xFe$ chain (Fig. 4b). The amount of field-induced splitting of the Kondo resonance is highly anisotropic, as is expected for spin systems with finite magnetic anisotropy and magnetic moments larger than ½ [21,23]. The impact of magnetic anisotropy can be maximized by replacing Mn atoms with Fe, e.g., switching from $Mn_3Fe$ to $MnFe_3$ (Fig. 4d). This substitution minimizes entanglement within the $m_T = \pm^1/_2$ ground states and suppresses the Kondo resonance.

This demonstrates that the emergent Kondo state can be tuned by the specific atomic composition of the spin chain. Such controllable creation of many-body states via engineering of the underlying quantum entanglement indicates that surface-adsorbed spin chains can serve as prototype systems for the exploration and simulation of correlated condensed-matter phases. In particular, it is now possible to access the transition from Fermi-liquid to non-Fermi liquid behaviour and to explore the exotic properties of the singular Fermi liquid associated with the underscreened Kondo phase. This approach provides a bridging step to bottom-up creation of real-world correlated-electron systems. Beyond testing prominent theoretical models such as quantum criticality in the two-impurity Kondo model[28] this method can in principle be generalized to a broader class of materials where electron-electron interaction can be tuned by structure and composition, such as heavy fermion compounds[29], or Mott insulators[30].

**METHODS**

**Experimental parameters:** All measurements used a low-temperature ultrahigh vacuum scanning tunnelling microscope (Unisoku USM 1300 3He) with base temperature of 0.5 K. A clean Cu(100) surface with large monatomically flat terraces was prepared by repeated Ar sputtering and annealing to 850 K. Then, a monolayer of copper nitride, $Cu_2N$, was formed as a decoupling layer by sputtering with $N_2$ at 1 kV and annealing to 600 K. Fe and Mn atoms were deposited onto the precooled surface by thermal evaporation of elemental Fe and Mn from effusion cells. All atom chains were created by vertical atom manipulation[17,18]. Prior to manipulation Fe and Mn atoms were identified using their characteristic conductance spectra[24]. Conductance spectra, $dI/dV(V)$, record the differential conductance of the tunnel junction as a function of sample bias detected by Lock-In detection of the tunnelling current caused by adding a modulation voltage of 72 $\mu V_{rms}$ amplitude at 691 Hz to the bias.

**Model spin Hamiltonian parameters:** The model applied to the spin chains uses an effective spin Hamiltonian to describe the spin states of the $Mn_xFe$. Conductance spectra are calculated by considering electron-spin scattering with tunnelling electrons (Supplementary Figure S2). The spin states of the MnFe dimers represented by their density matrices, $\rho$, are completely determined by the ratio $J/|D|$ (Fig. 2d). We quantify the degree of entanglement for the dimers (Fig. 2fg) by the entanglement entropy, $I(\rho)$. It is obtained by calculating the von Neumann mutual information of the Fe and Mn spin subsystems, $I(\rho) = S(\rho_{Fe}) + S(\rho_{Mn}) - S(\rho)$. $S$ is the von Neumann entropy and $\rho_{Fe}$ and $\rho_{Mn}$ are the reduced density matrices of the Fe and Mn spin obtained by partial trace over $\rho$. Since the spins considered here are larger than ½ we normalised $I(\rho)$ to the entanglement entropy of the maximally entangled state.

**Density function theory calculations:** Collinear DFT+U calculations were performed with the VASP code, using the PBE xc functional together with charging-energy corrections in the Dudarev et al. scheme (See Supplementary Section S2 for details on the calculation.). The d electrons for Fe and Mn had an onsite energy of $U$-$J$ = 1 eV and 4.0 eV respectively. For all chains we find the magnetic ground state of the $Mn_xFe$ chains to be close to $m_T = {}^1/_2$ due to antiferromagnetic ordering of the Fe and Mn magnetic moments (see Fig. 2c).

**ACKNOWLEDGEMENTS**

D.J.C., J.A.J.B., S.Y., S.R.P. and S.L. acknowledge Edgar Weckert and Helmut Dosch (Deutsches Elektronen-Synchrotron, Hamburg, Germany) for providing high-stability lab space. D.J.C. and J.A.J.B. acknowledge postdoctoral fellowship from the Alexander von Humboldt foundation. M.T. acknowledges support from the DFG-SFB 767. J.A.J.B. acknowledges support from the Natural Sciences and Engineering Research Council of Canada. N.L. and R.R. acknowledge support form Spanish MINECO (Grant No. MAT2012-38318-C03-02 with joint financing by FEDER Funds from the European Union). ICN2 acknowledges support from the Severo Ochoa Program (MINECO, Grant SEV-2013-0295).

**Figure 1:**

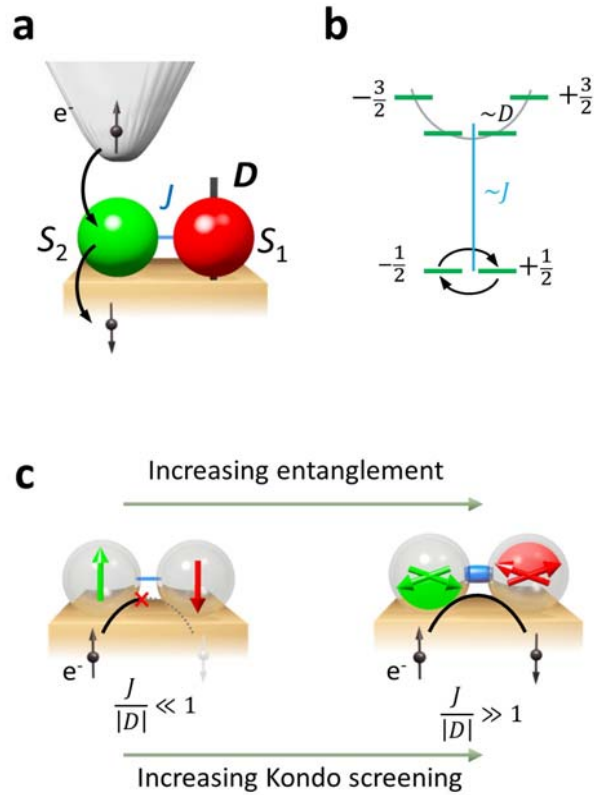

**Figure 1| Relation between Kondo screening and entanglement in atomically assembled spin chains. a.** Key properties for Kondo screening of the entangled spin dimer. Two atoms are coupled antiferromagnetically by Heisenberg interaction, *J* (blue). The magnetic moments of the atoms have spin magnitude $S_1$ and $S_2$ and differ by half-integer amount. Axial magnetic anisotropy, *D* (black), may be present. **b,** Spin state structure for a MnFe dimer (green lines). The doubly degenerate ground state with $m_T = \pm^1/_2$ is amenable to Kondo screening (black arrows). All excited states are split off by an energy proportional to *J* and modified by *D*. **c,** Relation between spin entanglement and Kondo screening: in the absence of entanglement, i.e. for $J/|D|\ll 1$, electron-spin scattering is forbidden and no Kondo screening can occur. When the spins $S_1$ and $S_2$ are strongly entangled, i.e. for $J/|D|\gg 1$, electron-spin scattering becomes possible and Kondo screening emerges.

**Figure 2:**

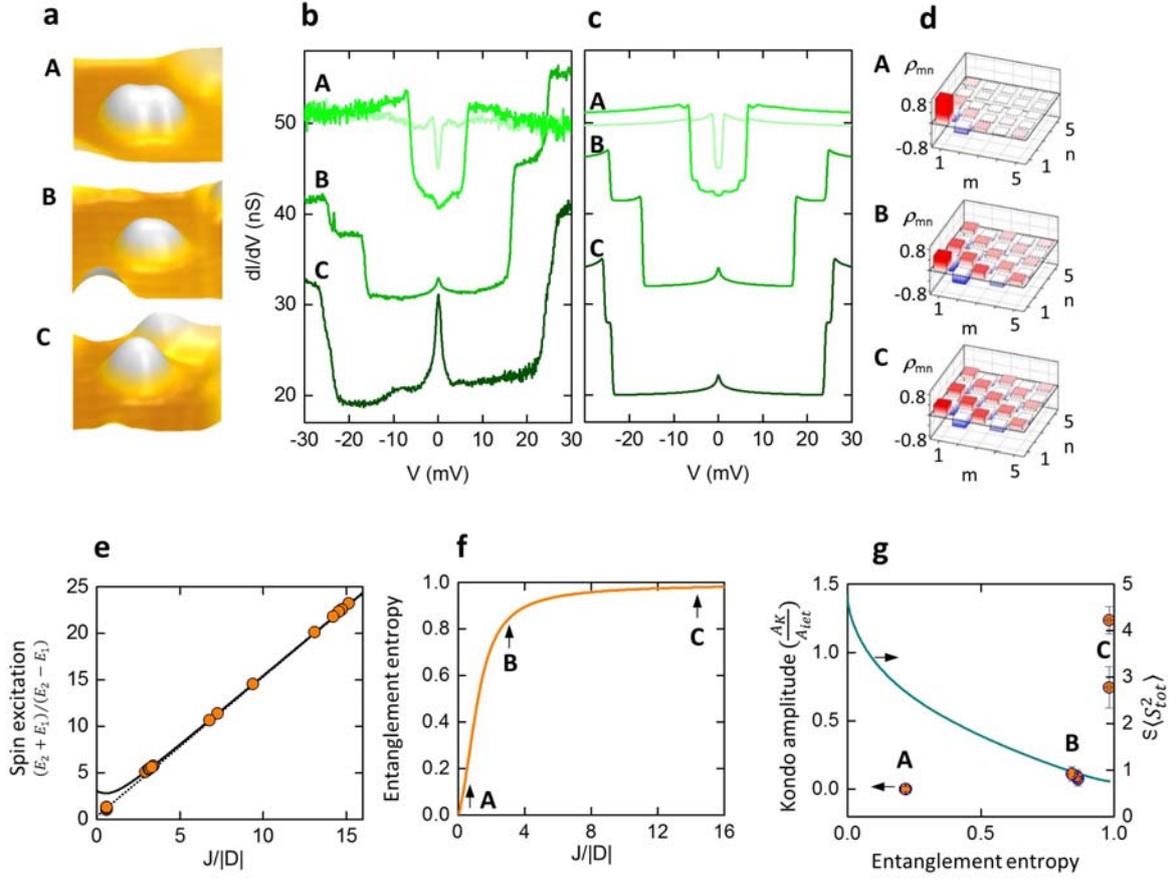

**Figure 2| Entanglement of MnFe dimers. a,** Constant-current topographs of MnFe dimers with separation of 0.72 nm (A), 0.36 nm (B) and a compact dimer (C). Image size (3.4 × 2.3) nm$^2$ shown as 3d rendering of height with color from low (orange) to high (white). **b,** Differential conductance spectra, d$I$/d$V$($V$), as function of sample bias, $V$. Spectra for dimer A measured on the Mn (light green) and Fe atom (green). Dimer B and C show one uniform spectrum and feature a Kondo resonance. **c,** Simulated d$I$/d$V$($V$) spectra for the dimers A, B, C using antiferromagnetic exchange interaction (A: $J$ = 0.6 meV, B: $J$ = 13.2 meV, C: $J$ = 16.5 meV) and axial magnetic anisotropy of the Fe atom (A: $D$ = 1.8 meV, B: $D$ = 3.9 meV, C: $D$ = 0.9 meV). The Kondo peak for dimer C is underestimated in our model as it cannot capture the strong-coupling limit of the Kondo effect. **d,** Reduced density matrices, $\rho_{mn}$, for the +1/2 ground state for spectra shown in **c**. Basis states contributing to $\rho_{mn}$ are 1: $|+5/2, -2\rangle$, 2: $|+3/2, -1\rangle$, 3: $|+1/2, 0\rangle$, 4: $|-1/2, +1\rangle$, 5: $|-3/2, +2\rangle$. **e,** Ratio of spin excitation energies, $(E_2+E_1)/(E_2-E_1)$, as a function of $J/|D|$ for all measured MnFe dimers (organge dots). Calculated (solid line) and approximated (dashed line) behaviour is based on our model Hamiltonian (see main text for details). **f,** Entanglement entropy as a function of $J/|D|$ (see Methods for definition of entanglement entropy). **g,** Relative Kondo peak amplitude, $A_K/A_{iet}$, and expectation value of the total spin $\langle S_T^2 \rangle$ as a function of entanglement entropy. The amplitudes of the Kondo resonance, $A_K$, and spin excitations, $A_{iet}$, were extracted by fitting the d$I$/d$V$(V) spectra (Supplementary Figure S6).

**Figure 3:**

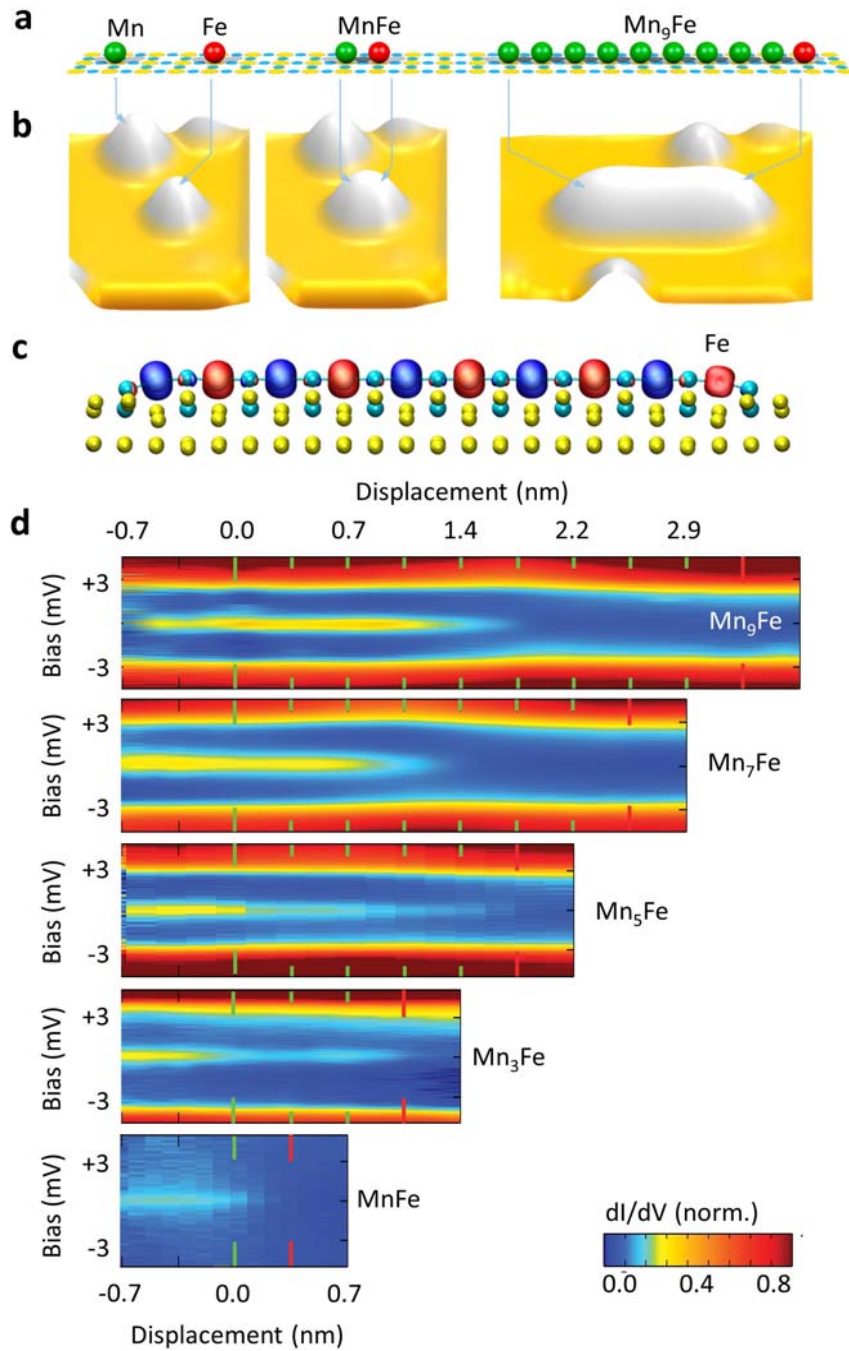

**Figure 3| Spatial maps of Kondo screening in Mn$_x$Fe spin chains. a,** Ball model of the construction of Mn$_x$Fe spin chains by adding Mn atoms (green) to an Fe atom (red). **b,** Constant current topographs of the chains shown in **a**. The Mn atom in the back provides atomic registry. **c,** Density difference between spin-up (blue) and spin-down (red) states over a Mn$_9$Fe chain. Spin density clouds have isosurface density of 0.05 e/Angstroms$^3$. Copper atoms (yellow) and Nitrogen atoms (cyan). (See Supplementary Section S2 for calculation details.) **d,** Maps of the differential conductance, d$I$/d$V$($V$,$d$), measured along the Mn$_x$Fe spin chains plotted colour-coded as a function of sample bias, $V$, and displacement, $d$, along the spin chain. Atom positions indicated by red (Fe) and green (Mn) dashes. $d$ = 0 set to the first Mn atom of each chain. The delocalized Kondo resonance is visible as horizontal yellow streak centred at 0 V bias.

**Figure 4:**

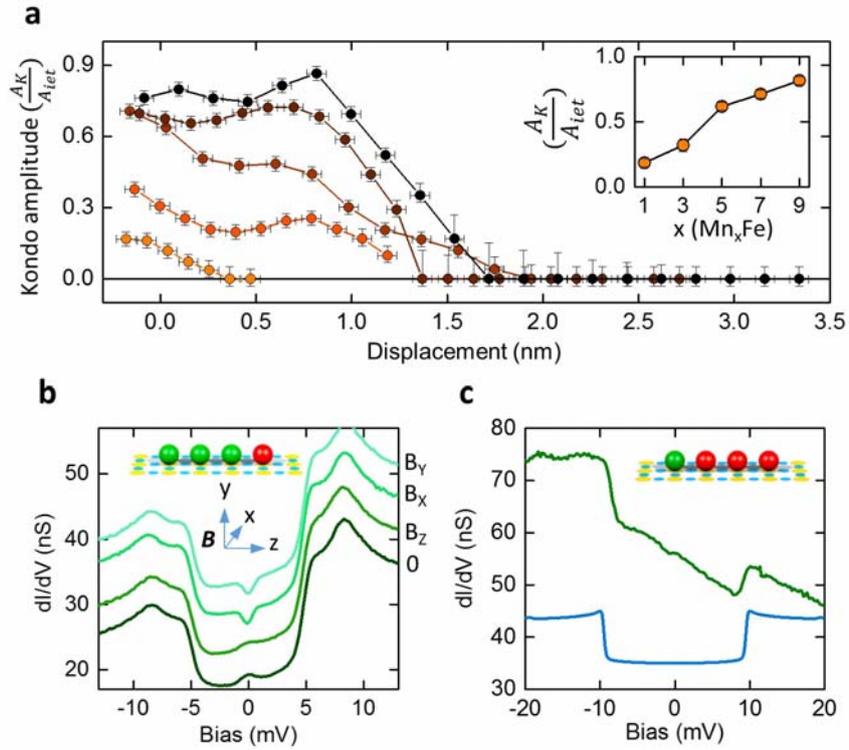

**Figure 4 | Evolution of Kondo screening in Mn$_x$Fe spin chains. a,** Spatial profile of the relative Kondo peak amplitude, $A_K/A_{iet}$, along each spin chain. Inset: Evolution of $A_K/A_{iet}$, with chain length at the first Mn atom. **b,** Conductance spectra, d$I$/d$V(V)$, showing the anisotropic magnetic-field dependent splitting of the Kondo resonance in Mn$_3$Fe at fields of 2T. Inset: ball model of the atomic structure of Mn$_3$Fe, Fe (red), Mn (green). Arrows indicate directions of the magnetic field. **c,** Measured (green) and simulated (blue) d$I$/d$V(V)$ spectra for MnFe$_3$ showing no Kondo resonance at 0 V. Parameters: $J$ = 8.8 meV, $J''$ = 11.0 meV (coupling between Fe and Fe), $D$ = 4.0 meV. Inset: ball model of Fe$_3$Mn structure.

SUPPLEMENTARY INFORMATION
# Entanglement-induced Kondo screening in atomic spin chains

Deung-Jang Choi[1,2,§], Roberto Robles[3], Shichao Yan[1,2], Jacob A. J. Burgess[1,2], Steffen Rolf-Pissarczyk[1,2], Jean-Pierre Gauyacq[4], Nicolás Lorente[5,6], Markus Ternes[2], Sebastian Loth[1,2,*]

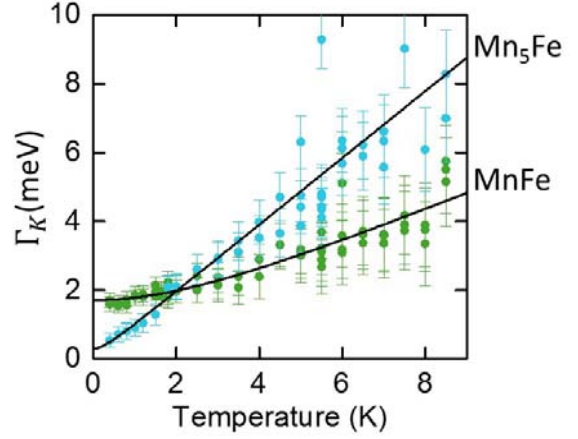

**Supplementary Figure S1 | Temperature-dependent broadening of Kondo resonances in $Mn_xFe$**. Half-width at half maximum of the Kondo resonances, $\Gamma_K$, determined by fitting the experimental spectra with inelastic tunneling steps and a Frota function (see Fig. S6) for MnFe (green), and $Mn_5Fe$ (blue). Black lines are fits of $\Gamma_K(T)$ to the function $\Gamma_K = \frac{1}{2}\sqrt{(2\Gamma_K^0)^2 + (3.5 k_B T)^2 + (\alpha k_B T)^2}$, which accounts for the thermal broadening of the tip's Fermi function and to the thermal broadening of the resonance [1-22]. For MnFe we obtain $\Gamma_K^0 = 1.68$ meV, $\alpha = 11.1$ and for $Mn_5Fe$ $\Gamma_K^0 = 0.29$ meV and $\alpha = 22.3$. The large values for $\alpha$ confirm the correlated nature of the resonances and are consistent with Kondo-screened spin systems in the weak coupling regime [4] where the characteristic Kondo temperature is below the lowest measurement temperature of T = 0.5 K.

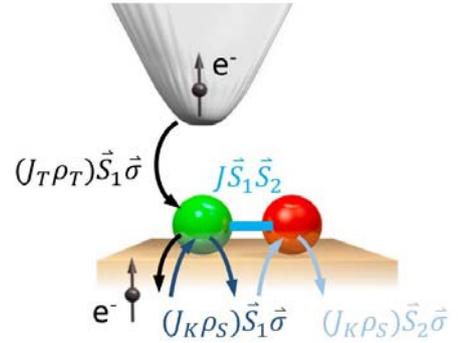

**Supplementary Figure S2 | Schematic transport model.** The $dI/dV(V)$ spectra were simulated using a perturbative approach which accounts for spin-flip scattering processes up to 3rd order in the interaction matrix elements. The transition probability $W_{i \to f}^a$ for an electron to tunnel from the tip placed above the $a$-th spin to the sample (or from sample to the tip) by simultaneously changing the quantum state of the spin system between the initial, $i$, and final, $f$, state is in this model given as:

$$W_{i \to f}^a \propto (J_T \rho_T)^2 \left( \left| M_{i \to f}^a \right|^2 + \sum_{b=1}^n J_K \rho_S \sum_m \left( \frac{M_{f \to i}^a M_{m \to f}^b M_{i \to f}^a}{\varepsilon_i - \varepsilon_m} + c.c. \right) \right) \delta(\varepsilon_i - \varepsilon_f)$$

.
Here, $M_{i \to j}^a = \langle \psi_j, \sigma_j | \hat{S}^a \cdot \hat{\sigma} | \psi_i, \sigma_i \rangle$ is the scattering matrix element of the Kondo-like interaction between the combined state vector $|\psi_i, \sigma_i\rangle$ and $|\psi_j, \sigma_j\rangle$ with $\psi_i$ as the eigenstate of the localized spin system and $\sigma_i$ as the wave vector of the interacting electron. $\hat{S}^a$ is the total spin operator acting on the $a$-th spin of the coupled structure, and $\hat{\sigma}$ the standard Pauli matrices. The delta distribution in the equation accounts for energy conservation between the initial state energy, $\varepsilon_i$, and the final state energy, $\varepsilon_f$. The first term accounts for spin-flip excitation which produces conductance steps in the spectra. The second term leads to logarithmic peaks at the intermediate energy, $\varepsilon_m$, and scales with the coupling strength to the substrate, $J_K\rho_S$. The scattering strength of tunneling electrons from the tip, $J_T\rho_S$, is scaled to match the average differential conductance of the experimental spectra between 20 nS to 30 nS. $J_K\rho_0$, was set to −0.02. This value was chosen to match the intensity of the Kondo resonance of the MnFe dimer with 0.36 nm atomic separation (Fig. 2c, dimer B) and left constant throughout this work to allow comparison between different spin chains. Note, that this model cannot cover strong correlations because it neglects higher order effects and may therefore underestimate the Kondo resonance's amplitude for maximally entangled spin chains.

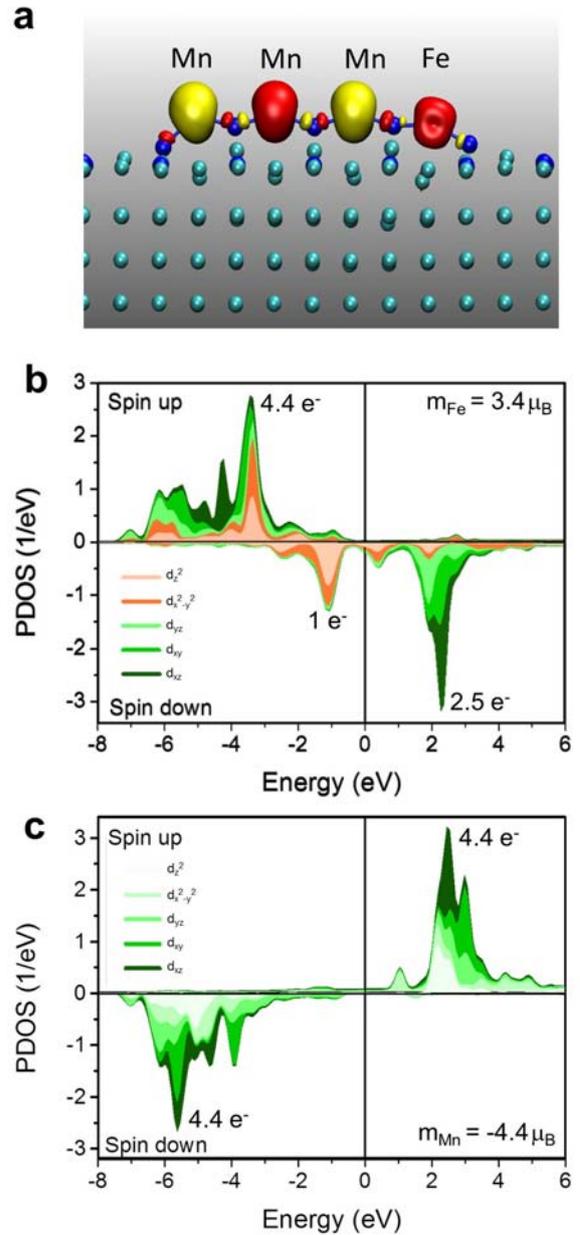

**Supplementary Figure S3 | Density functional calculations on Mn₃Fe chain. a,** Spin density for the Mn₃Fe chain on Cu$_2$N/Cu (100) for an isovalue of 0.05 electrons/Angstroms$^3$. Red is for spin up densities and yellow for spin down. The left atom of the chain corresponds to the Fe atom. The total magnetic moment of the Fe atom corresponds to 3.4$\mu_B$. For the Mn atoms it is -4.4$\mu_B$ for the atom closest to the Fe atom, +4.4$\mu_B$ for the central atom and -4.4$\mu_B$ for the end Mn atom. The total spin of the chain is then almost $1/2$. The nitrogen atoms (dark blue) become part of the chain and show alternating spins facing the transition metal atoms in agreement with super-exchange being the origin of the antiferromagnetic spin ordering. **b.** Integrated projected density of states (PDOS) on selected $d$-atomic orbitals of the Fe atom of the Mn₃Fe chain. The Fe atom's electronic structure shows occupied states with both spin up and spin down character. **c,** PDOS for the Mn atom adjacent to Fe. The Mn atom's spin polarization is stronger and contrary to Fe, the $d_{z^2}$ orbital is only partially filled. The PDOS for the other Mn atoms is similar. It is worth noting that the $d$ electrons of Mn carry the full magnetic moment and are separated from the Fermi energy (Energy=0) by several electron Volts. Therefore they cannot be Kondo-screened individually.

**Supplementary Figure S4 | Full series of $Mn_xFe$ (x = 1, ..., 10) spin chains.** Conductance spectra, dI/dV(V) for each chain were measured on the edge Mn atom that is furthest from the Fe atom. Every chain with an odd number of Mn atoms (x = 1, 3, 5, 7, 9) shows a Kondo resonance at zero sample bias and an excitation gap that shrinks with increasing chain length. Every chain with even number of Mn atoms (x = 2, 4, 6, 8, 10) shows a gapped spectrum; no Kondo resonance can be found. These spectra fit well to the same model Hamiltonian as the $Mn_xFe$ (x odd) chains described in the main text. For an even number of Mn atoms the spin chain resides in a ground state with $m_T = \pm 2$ which cannot be Kondo-screened by a single-electron interaction. Temperature for all spectra, $T = 0.5$ K, and magnetic field, $B = 0$ T. Spectra offset vertically for clarity.

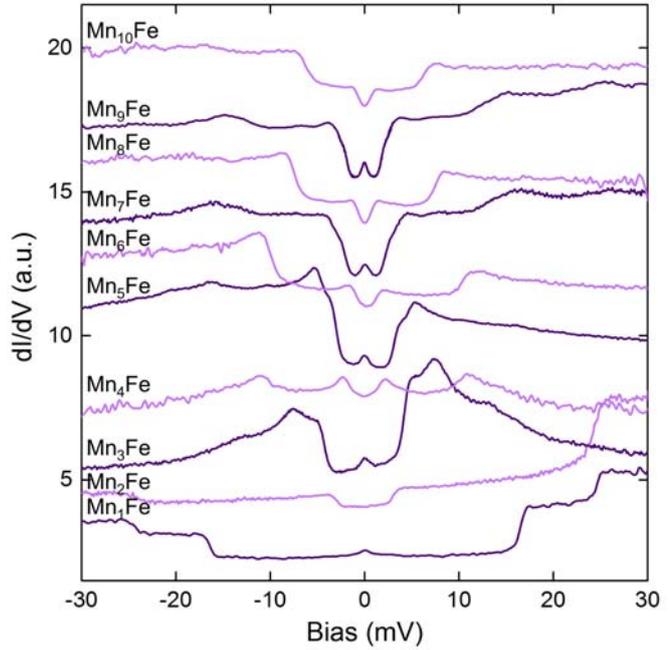

**Supplementary Figure S5 | Evolution of spin excitation energies for $Mn_xFe$ chains.** All $Mn_xFe$ chains with odd number of Mn atoms show a similar dI/dV(V) spectrum. A Kondo resonance is present at 0 V and two low-energy spin excitations are visible. Orange dots: Measured spin excitation energies determined by the full fit to the dI/dV(V) spectra for each $Mn_xFe$ chain (Fig. S6). As sketched in Fig. 1b and outlined in the main text, these two spin excitations enable fitting the spin Hamiltonian model quantitatively to the experimental data. Blue dots: Calculated spin excitation energies resulting from our spin Hamiltonian model by constraining the fit to use the same set of parameters for all chains. Fit results are: $J = 13.3$ meV $\pm 1.0$ meV (spin coupling between Fe and Mn), $J' = 4.2$ meV$\pm 0.8$ meV (spin coupling between Mn and Mn), $D = 4.0$ meV $\pm 0.4$ meV (magnetic anisotropy of the Fe atom).

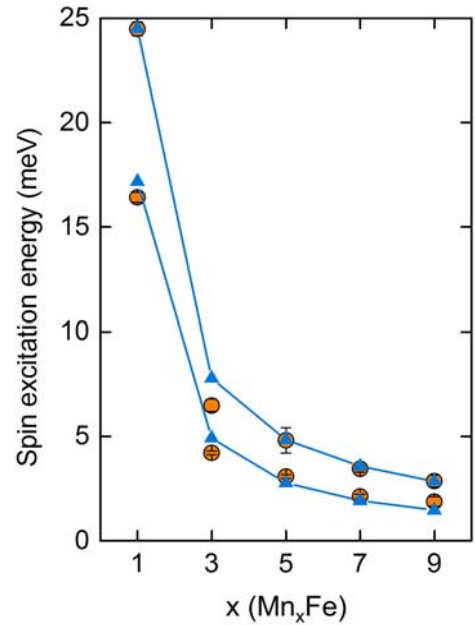

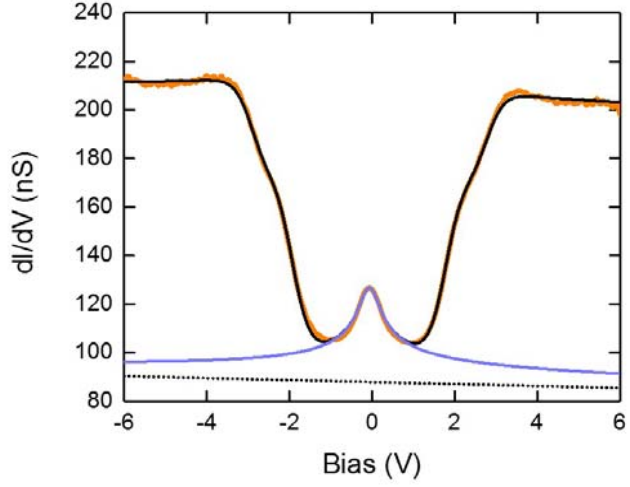

**Supplementary Figure S6 | Fitting of experimental conductance spectra for Mn$_x$Fe chains.** All conductance spectra of Mn$_x$Fe chains with an odd number of Mn atoms show two prominent spin excitations and a Kondo resonance. Graph shows exemplary fit for one spectrum measured on Mn$_9$Fe at T = 0.5 K and B = 0 T. The measured spectrum (orange line) is fit by least-squares optimization of the function (black)

$$\sigma(V) = A_1^- \theta(-V_1) + A_1^+ \theta(V_1)$$
$$+ A_2^- \theta(-V_2) + A_2^+ \theta(V_2)$$
$$+ A_K f(V, \Gamma) + \sigma' V + \sigma_0$$

The two spin excitations are described by temperature-broadened step functions [5], $\theta(V)$, and the Kondo resonance as a Frota function [6] (blue), $f(V,\Gamma)$, with a half-width at half maximum of $\Gamma_K$. $A_{1,2}^{+,-}$ are the amplitudes of the first and second spin excitation at positive and negative energy. $V_1$ and $V_2$ are the voltage positions of each spin excitation. $A_K$ is the amplitude of the Kondo resonance and $\Gamma_K$ is its width. $\sigma'$ and $\sigma_0$ account for a sloped background conductance (dotted line). We define the average spin excitation step height as $A_{iet} = \frac{1}{4}\sum_{i,p} A_i^p$ and the relative Kondo peak height as $A_K/A_{iet}$.

**Supplementary section S1: Underscreened Kondo phase in Mn$_x$Fe chains.**

Evidence of an incomplete screening of the magnetic moment on the Mn$_x$Fe spin chains is found in the highly anisotropic splitting of the Kondo resonance with magnetic field (main text, Fig. 4b). Magnetic fields applied perpendicular to the easy-axis of the Fe atom (the direction parallel to the chain) induce a rapid decrease in resonance amplitude and a splitting can be observed for magnetic fields in excess of 1 T. In contrast, magnetic fields applied parallel to the Fe atom's easy axis induce no splitting up to the highest field we could apply (2 T).

The Kondo resonance observed in the Mn$_x$Fe chains results from electron-spin scattering at several high-spin atoms, i.e., Fe with spin magnitude $S_{Fe} = 2$ and Mn with spin magnitude $S_{Mn} = 5/2$. Contrary to the typically considered Kondo scattering at a single magnetic impurity with spin magnitude $1/2$, the Mn$_x$Fe chains can be in a situation where the degenerate ground state doublet is a superposition with spin states that have a total spin in excess of $½$. Indeed, the mean value of $\langle S_T^2 \rangle$ (the expectation value of the total spin operator for the ground state) for all Mn$_x$Fe chains is larger than that of a pure $S = ½$ macrospin.

The anisotropic splitting in conjunction with strong temperature-dependent broadening of the resonance (Fig. S1) are direct evidence for a departure from the fully-screened and strongly-coupled Kondo phase. A similarly anisotropic dependence on the orientation of the magnetic field was found for Co atoms on Cu$_2$N/Cu(100) [21]. Calculations by Zitko et al [8] show that in the presence of easy-axis magnetic anisotropy, only some $S_z$ components of the total spin are Kondo-screened. The remaining magnetic interactions lead to an underscreened Kondo phase.

**Supplementary Section S2: Details on the Density Function calculations of Mn$_x$Fe chains.**

Collinear DFT+U calculations were performed with the VASP code [9, 10], using the PBE xc functional [11] together with charging-energy corrections in the Dudarev *et al.* [12] scheme. The calculations use a slab geometry with 4 layers and 36 Cu atoms per layer for the Mn$_3$Fe chains. The *k*-point sampling was $4 \times 2 \times 1$. For larger chains we increased the supercell size, keeping three-lattice-parameter distance (10.95 Angstroms, the PBE lattice parameter for Cu is 3.65 Angstroms in the present calculations) between chains along the chain direction and across the chains. The energy cutoff was kept at 300 eV. The d electrons for Fe and Mn had an onsite energy of *U-J* = 1 eV and 4.0 eV respectively. The chain and the first two surface layers were relaxed until forces where below 0.01 eV/Angstrom.

For all chains we find an $S_z$=1/2 ground state due to the antiferromagnetic (AF) ordering of the Fe and Mn magnetic moments. These moments are slightly smaller than those of free atoms, as has already been found in extensive calculations of Mn chains on Cu$_2$N [13], but the AF ordering is maintained independently of the length of the chain as can be seen in Fig. 3c and Supplementary Figure S3 for Mn$_9$Fe and Mn$_3$Fe respectively. Moreover, the N atoms taking part in the foundation of the chains present displaced spin densities, albeit with a zero local magnetic moment. The polarised spin densities alternate on both sides of the N atoms and show the spin structure of the atomic mediated super-exchange interaction between local moments [6]. In the super-exchange interaction, the non-magnetic species share two opposite spins with neighbouring magnetic atoms. Since the two spins are opposite, the induced interaction is AF.